\documentclass[prl,reprint,showpacs,superscriptaddress,aps]{revtex4-1}

\usepackage{graphicx}
\usepackage{bm}
\usepackage{color}
\usepackage[USenglish]{babel}
\usepackage{amsmath}
\usepackage{braket}
\usepackage{multirow}
\usepackage{arydshln}
\usepackage{hyperref}
\usepackage{etoolbox}
\hypersetup{
	colorlinks=true,
	linkcolor=blue,
	citecolor=blue,
	urlcolor=blue
}
\begin{document}
	
	
	\title{Spin-rotation coupling in $p$-wave Feshbach resonances}
	\author{Bing Zhu}
	\email{bzhu@physi.uni-heidelberg.de}
	\affiliation{Physikalisches Institut, Universit\"at Heidelberg, Im Neuenheimer Feld 226, 69120 Heidelberg, Germany}
	\affiliation{Hefei National Laboratory for Physical Sciences at the Microscale and Department of Modern Physics, and CAS Center for Excellence and Synergetic Innovation Center in Quantum Information and Quantum Physics, University of Science and Technology of China, Hefei 230026, China}

	\author{Stephan H\"afner}
	\author{Binh Tran}
	\author{Manuel Gerken}
	\author{Juris Ulmanis}
	\affiliation{Physikalisches Institut, Universit\"at Heidelberg, Im Neuenheimer Feld 226, 69120 Heidelberg, Germany}
	
	\author{Eberhard Tiemann}
	\email{tiemann@iqo.uni-hannover.de}
	\affiliation{Institut f\"ur Quantenoptik, Leibniz Universit\"at Hannover, Welfengarten 1, 30167 Hannover, Germany}
	
	\author{Matthias Weidem\"uller}
	\email{weidemueller@uni-heidelberg.de}
	\affiliation{Physikalisches Institut, Universit\"at Heidelberg, Im Neuenheimer Feld 226, 69120 Heidelberg, Germany}
	\affiliation{Hefei National Laboratory for Physical Sciences at the Microscale and Department of Modern Physics, and CAS Center for Excellence and Synergetic Innovation Center in Quantum Information and Quantum Physics, University of Science and Technology of China, Hefei 230026, China}

	\date{\today}
	
\begin{abstract}
We report evidence for spin-rotation coupling in $p$-wave Feshbach resonances in an ultracold mixture of fermionic $^6$Li and bosonic $^{133}$Cs lifting the commonly observed degeneracy of states with equal absolute value of orbital-angular-momentum projection on the external magnetic field. By employing magnetic field dependent atom-loss spectroscopy we find triplet structures in $p$-wave resonances. Comparison with coupled-channel calculations, including contributions from both spin-spin and spin-rotation interactions, yields a spin-rotation coupling parameter $|\gamma|=0.566(50)\times10^{-3}$. Our findings highlight the potential of Feshbach resonances in revealing subtle molecular couplings and providing precise information on electronic and nuclear wavefunctions, especially at short internuclear distance. The existence of a non-negligible spin-rotation splitting may have consequences for future  classifications of $p$-wave superfluid phases in spin-polarized fermions.
\end{abstract}

\maketitle
	
The notion of spin-rotation (\emph{sr}) coupling in diatomic molecules dates back to the early days of quantum mechanics, first being introduced by Friedrich Hund in 1927 to explain the multiplet splitting of rotational levels in nonsinglet molecular states \cite{Hund1927}. It is caused by the interaction of the electron's magnetic moment with the magnetic fields created either directly by the rotating nuclear charges (first-order contribution) \cite{Kramers1929}, or by the distorted electronic orbits due to the nuclear pair rotation inducing an effective electronic spin-orbit interaction (second-order contribution) even for states with vanishing electronic angular momentum, like, e.g., $^3\Sigma$ molecules \cite{Vleck1929}. Another important effect determining the fine structure of rotational levels in molecules with unpaired electrons is the spin-spin (\emph{ss}) interaction, which also arises from both a direct interaction (magnetic dipole-dipole interaction (mDDI) in this case) and a second-order contribution (electronic spin-orbit coupling) \cite{Brown2003}. The first-order \emph{sr} interaction involves a magnetic dipole moment induced by the rotation of the nuclei, while the mDDI is directly proportional to the electron's magnetic moment. Thus, the relative strength of \emph{sr} interaction relative to \emph{ss} scales as the mass ratio $\sim m_e/m_p \approx 5\times10^{-4}$. For the second-order contribution, the relative scaling can be estimated as $B/A$, where $B$ is the molecular rotational energy and $A$ \emph{so} coupling constant \cite{Tinkham1955}. Generally, except for very light molecules like H$_2$, the second-order effects dominate in deeply bound non-singlet molecules. 


One may now wonder whether these interactions have measurable consequences in very weakly bound molecular states as encountered in magnetically tunable Feshbach resonances (FRs) in ultracold collisions between neutral atoms \cite{Chin2010}. The couplings due the electron's spin magnetic moment are only present in collision channels involving the $a^3\Sigma^{+}$ triplet molecular ground state. In fact, \emph{ss} interaction gives rise to a well-established doublet (triplet) splitting in $p$-wave ($d$-wave) FRs \cite{Ticknor2004, Cui2017, Yao2019}, with an observed degeneracy according to the projection $|m_N|$ of pair rotation angular momentum $N$ along the external magnetic field. As an example related to our experiments, the \emph{ss} splitting of $p$-wave FRs in $^6$Li-$^{133}$Cs system was determined to be $\sim400$~mG \cite{Repp2013}. Due to the large internuclear distance, the rotational energy $B$ is below 1~GHz (in units of Planck's constant), while the spin-orbit coupling constant $A$ approaches the atomic value. Thus, the ratio $B/A$ ranges between $10^{-2}$ for $^6$Li and $10^{-5}$ for $^{133}$Cs, suggesting that not only first-order, but also second-order \emph{sr} coupling can be safely neglected as they would lead to additional splittings far beyond the magnetic field resolution in a typical ultracold-atom experiment exploring FRs (for details on the \emph{sr} coupling estimation, see \cite{Zhu2019}).

\begin{figure*}[t]
	\centering
	\includegraphics[width=\textwidth]{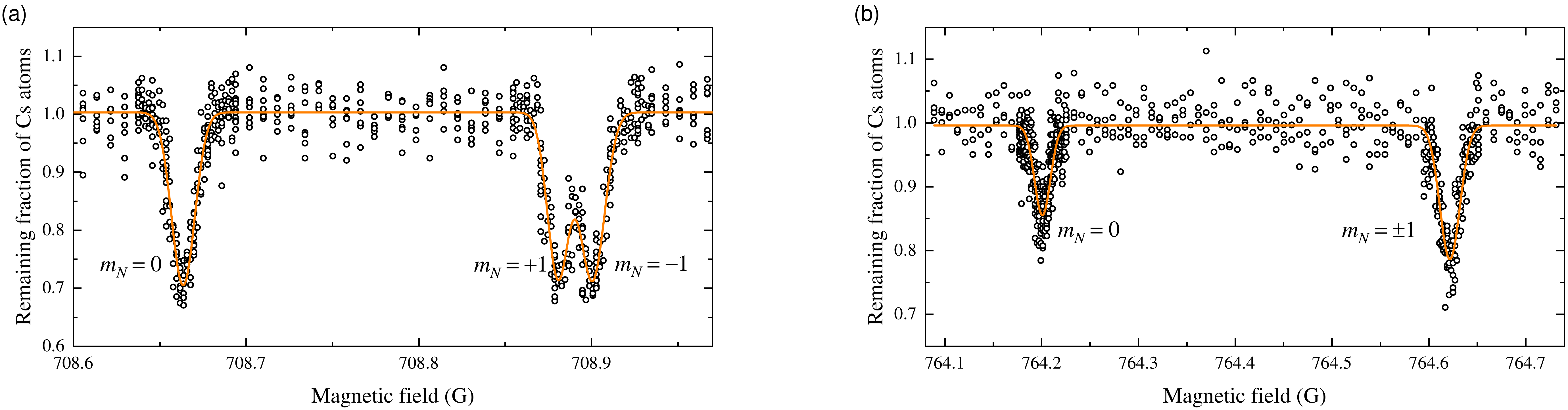}
	\caption{Li-Cs $p$-wave Feshbach resonances in the Li$\left|1/2,-1/2\right\rangle \oplus $Cs$\left|3,3\right\rangle$ channel observed in the remaining fraction of Cs atoms after a fixed holding time. (a) Triplet structure observed in a $p$-wave FR close to 709~G (holding time 1~s). (b) Doublet structure observed in a $p$-wave FR close to 764~G (holding time 5~s). Loss of the Li atoms shows identical resonance features. The magnetic field is randomly sampled and we reduce the field step size from 8~mG to 2~mG around the loss features. The solid lines are fits of a multi-peak Gaussian function. The assignment of $m_l$ is from cc calculation assuming a positive $\gamma$ (see main text).
	}
	\label{fig:pFRCs}
\end{figure*}

In this letter we present high-precision atom-loss spectroscopy on magnetically induced $p$-wave FRs in an ultracold mixture of $^6$Li and $^{133}$Cs atoms. Around magnetic fields of 658 G, 663 G, and 709 G, we find a splitting of the $p$-wave FRs into a triplet structure, resolving all three projections $m_N=-1,0,+1$ of the rotational angular momentum $N=1$. Based on a full coupled-channels (cc) calculation the observed splitting can be attributed to the \emph{ss} and \emph{sr} interactions corresponding to energy scales of $0.01 B$ and $0.0005 B$, respectively ($B = h \times 288$ MHz is the rotational energy of the least bound vibrational state of the LiCs triplet ground state). Although the wavefunction of such a weakly bound molecular state is centered at large internuclear distances, it turns out that second-order effects involving spin-orbit coupling, which are significant at small internuclear distances only, need to be included to fully describe both the observed \emph{ss} and \emph{sr} splittings \cite{Zhu2019}. Our findings demonstrate once again that high-resolution Feshbach spectroscopy exquisitely provide accurate information on the various angular-momentum couplings as well as the nuclear and electron wavefunctions in a diatomic molecule, especially at short and intermediate internuclear distances.    

Our experimental procedure is similar to the one presented in Refs. \cite{Pires2014,Ulmanis2015}. We prepare an optically trapped ultracold mixture of fermionic $^6$Li atoms and bosonic $^{133}$Cs atoms by means of standard laser-cooling techniques. The Cs atoms are optically pumped and spin-polarized in their absolute ground state $\left|f=3,m_f=3\right\rangle$ during degenerate Raman sideband cooling \cite{Kerman2000} and then loaded into a crossed optical dipole trap. Here, $f$ and $m_f$ refer to the total angular momentum and its projection on the external field axis. Subsequently the spin-balanced Li atoms in the $\left|f=1/2\right\rangle$ manifold are trapped in a second, cigar-shaped optical dipole trap, located approx. 1~mm away from the Cs cloud. The two atomic species are evaporatively cooled for 3~s separately at a magnetic field of 818~G and finally combined in the dipole trap at a magnetic field of 907~G. The Li atoms are spin-polarized by removing one of the spin states with a short resonant light pulse. Finally, $5\times 10^4$ Cs and $3\times 10^4$ Li atoms in their respective spin states are trapped at a temperature of 430~nK for the $p$-wave measurements. 

\begin{table*}[t]
	\caption{Multiplet splitting in Li-Cs $p$-wave FRs observed at the magnetic field $B^e_{0}$. $\delta_{ss}^{e}$ and $\delta_{sr}^{e}$ are the measured splittings due to the \emph{ss} and \emph{sr} interaction, respectively. The numbers in brackets give the uncertainties in the determination of $B^e_{0}$, $\delta_{ss}^{e}$, and $\delta_{sr}^{e}$. The systematic uncertainty is shown for $B^e_{0}$ in the second bracket. The corresponding theoretical values $\delta_{ss}^{cc}$ and $\delta_{sr}^{cc}$ are obtained from the cc scattering calculation for a relative kinetic energy corresponding to the temperature of 430~nK as in the experiment. $\Delta E_{ss}$ and $\Delta E_{sr}$ are the calculated energy differences by the \emph{ss} and \emph{sr} interactions.}
	\label{tab:pwaves}
	\begin{ruledtabular}
		\begin{tabular}{lccccccc}
			Entrance channel & $B^e_0$ (G) & $\delta_{ss}^{e}$ (mG) & $\delta_{ss}^{cc}$ (mG) & $\Delta E_{ss}$ ($h\times$MHz) & $\delta_{sr}^{e}$ (mG) & $\delta_{sr}^{cc}$ (mG) & $\Delta E_{sr}$ ($h\times$MHz)\\\hline
			Li$\left|1/2,+1/2\right\rangle$	&662.822(3)(16) & 224(4) & 228.5 & 0.70 & 20(6) & 17 & 0.052\\
			$\oplus$ Cs$\left|3,+3\right\rangle$ &713.632(4)(16) & 422(5) & 419.5 & 1.19  & -\footnote{No splitting is observed.} & 9 & 0.026\\\hline
			
			Li$\left|1/2,-1/2\right\rangle$ &658.080(10)(16) & 75(12) & 81.5 & 0.27 & 24(14) & 25 & 0.082\\
			$\oplus$ Cs$\left|3,+3\right\rangle$ &708.663(3)(16) & 228(4) & 232.5 & 0.71 & 20(4) & 17 & 0.052\\
			&764.201(1)(16) & 421(1) & 423 & 1.20 & -\footnotemark[1] & 10 & 0.028\\
		\end{tabular}
	\end{ruledtabular}	
\end{table*}

FRs are identified as simultaneously enhanced atom losses in both Li and Cs after an optimized holding time. In Fig.~\ref{fig:pFRCs} we show typical loss features in the measurements of the remaining fraction of Cs atoms at randomly sampled magnetic fields. The loss of Li atoms showing identical resonance features are presented in Ref. \cite{Zhu2019}. We assign each loss peak with quantum number $m_N$, according to the cc calculation. The resonance positions $B^e_{m_N}$ and widths $\sigma_{m_N}$ are extracted by fitting a (multi-)peak Gaussian function. Figures~\hyperref[fig:pFRCs]{1(a)} and \hyperref[fig:pFRCs]{1(b)} show two $p$-wave FRs close to 709~G and 764~G with a triplet and doublet structure, respectively. In total, the triplet structure is observed in three $p$-wave FRs close to 658~G, 663~G, and 709~G, while it is not resolved for the 714~G and 764~G FRs (see Ref. \cite{Zhu2019}). This observation is consistent with the cc modeling (see below), which predicts splittings below the experimental resolution for the later two resonances. The experimental doublet splittings $\delta^e_{ss}=(B^e_{+1}+B^e_{-1})/2-B^e_{0}$ and triplet splittings $\delta^e_{sr}=B^e_{-1}-B^e_{+1}$ are summarized in Table \ref{tab:pwaves}.

The observed widths of all the $p$-wave FRs are around 10~mG. To resolve such narrow loss features, we refine the step size of the magnetic field from 8~mG to 2~mG and increase the number of repetitions at magnetic fields around the resonances. However, the observed widths are much larger than the calculated widths of the two-body collision rate peaks ($\ll$ 1~mG) deduced from the cc calculation and have thus to be attributed to the thermal energy distribution of the atoms, magnetic field variations over the size of the cloud, and short-term field fluctuations. 

Measurements at higher ensemble temperatures show a broadening of the loss features and a shift of the loss centers towards higher fields. The temperature-dependence of $B_{0}^e$ and $\delta^e_{sr}$ is shown for the 663-G resonance in Ref. \cite{Zhu2019}. The broadening, due to an increased thermal energy distribution, then leads to an unresolvable $m_N=\pm 1$ splitting. Therefore, it is essential to measure at ultralow temperatures to resolve the splitting. On the other hand, at lower temperatures, the rotational barrier suppresses the $p$-wave scattering cross section $\sigma_p\propto T^2$  \cite{DeMarco1999}, resulting in loss being dominated by the finite $s$-wave cross-section, single species loss mechanisms such as Cs three-body recombination, or the lifetime of the sample in the optical dipole trap. The experimental parameters were chosen to ensure that the splitting is well resolved while maintaining a sufficiently large interspecies loss rate.

The \emph{ss} interaction is described by the following operator separating the spatial and spin degree of freedoms \cite{Ticknor2004}
\begin{equation}
\begin{aligned}
H_{ss}&= 4\lambda(R) \sum\limits_{q=-2}^{2} (-1)^q(\hat{\mathbf{R}}\otimes\hat{\mathbf{R}})^{(2)}_{-q}(\mathbf{s}_1\otimes\mathbf{s}_2)^{(2)}_q \, ,\\
\end{aligned}
\label{eq:HamiltonianSS}
\end{equation}
where $\hat{\mathbf{R}}$ is the unit vector along the internuclear axis and $\mathbf{s}_i$ ($i=1,2$) are the atomic electron spins. $(\hat{\mathbf{R}}\otimes\hat{\mathbf{R}})^{(2)}_{-q}$ and $(\mathbf{s}_1\otimes\mathbf{s}_2)^{(2)}_q$ are rank-2 irreducible tensors formed from two vectors \cite{Sakurai2010} acting on the partial-wave components $\ket{N,m_N}$ and the spin states $\ket{S,S_z}$, respectively.  Here $S_z$ is the projection of the total electron spin $S$ onto the external magnetic field. Diagonal parts $\braket{1,m_N|(\hat{\mathbf{R}}\otimes\hat{\mathbf{R}})^{(2)}_{0}|1,m_N}$ account for the doublet splitting, e.g. $\frac{2}{5}\sqrt{\frac{2}{3}}$ for  $m_N=0$ and $-\frac{1}{5}\sqrt{\frac{2}{3}}$ for $m_N=\pm1$, which leads to an energy difference of
\begin{equation}
\begin{aligned}
\Delta E_{ss}&= \frac{2}{5}\braket{\lambda(R)}\braket{3S_z^2-S^2-1}
\end{aligned}
\label{eq:EnergSS}
\end{equation}
between $m_N=0$ and $m_N=\pm1$ components. Here, $\braket{\lambda(R)}$ and $\braket{3S_z^2-S^2-1}$ are calculated by the spatial and spin parts of the close-channel wave-function. Off-diagonal parts in $m_N$ ($q\neq0$ in Eq. \ref{eq:HamiltonianSS}) are too small at high magnetic fields to cause a significant splitting of the $m_N=\pm1$ components (see Ref. \cite{Park2012} at low magnetic fields).

In the cc calculation, the function $\lambda(R)$ in atomic units 
$$\lambda(R)=-\frac{3}{4}\alpha^2 \left(\frac{1}{R^3}+a_{\mathrm{so1}} \exp(-b_1 R)+ a_{\mathrm{so2}} \exp(-b_2 R) \right)$$
includes the mDDI \cite{Stoof1988,Moerdijk1995} with a $1/R^3$ functional form and the second-order \emph{so} interactions with biexponential R-functions \cite{Mies1996,Kotochigova2000}, dominating the \emph{ss} interaction at large and small $R$, respectively. $\alpha$ is the fine-structure constant and $a_{\mathrm{soi}}$ and $b_{i}$ are adjustable parameters. By fitting the observed doublet splittings $\delta^e_{ss}$ to a full cc model (see Table \ref{tab:pwaves}), we determine the parameters in the function above to be: $a_{\mathrm{so1}}=-1.99167$, $b_1=0.7$, and $a_{\mathrm{so2}}=-0.012380$, $b_2=0.28$. The observed \emph{ss} splittings correspond to $\braket{\lambda(R)}\sim$2.5~MHz (in units of h), about 0.01 of the rotational energy $B=\braket{\hbar^2/(2\mu R^2)}\sim 288$~MHz with the reduced mass $\mu$. The ratio $\lambda/B$ is about $10\sim100$ in deeply bound $^3\Sigma$ molecules \cite{Brown2003}.  

To model the observed triplet structures in the $p$-wave FRs, we include the \emph{sr} coupling in the Hamiltonian taking the form
\begin{equation}
\begin{aligned}
H_{sr}&=\frac{\gamma}{2\mu R^2} \mathbf{S} \cdot \mathbf{N}\\
\end{aligned}
\label{eq:HamiltonianSR}
\end{equation}
with the dimensionless \emph{sr} coupling parameter $\gamma$. Here, $\mathbf{S}$ and $\mathbf{N}$ are the total electron spin and molecular rotation angular momentum, respectively. At high magnetic fields both $N$ and $m_N$ are nearly conserved and only the diagonal matrix elements in $m_N$ of Eq. \eqref{eq:HamiltonianSR} contribute significantly. The diagonal terms read $\gamma\frac{\hbar^2}{2\mu R^2}S_z m_N$ which leads to an energy splitting of 
\begin{equation}
\begin{aligned}
\Delta E_{sr}=2\gamma B\braket{S_z} \, 
\end{aligned}
\label{eq:EnergySR}
\end{equation}
between $m_N=+1$ and $m_N=-1$ component, where $S_z$ is the projection of $\mathbf{S}$ along the external magnetic fields and $\braket{S_z}$ is its expectation value in the closed channel. This energy difference will be translated into the observed splitting $\delta^e_{sr}$ by the differential magnetic moments between closed and open channel.

As compared in Table~\ref{tab:pwaves}, within the experimental uncertainties both the observed doublet (\emph{ss}) and triplet (\emph{sr}) splittings are reproduced very well by the cc calculation including the Hamiltonians \eqref{eq:HamiltonianSS} and \eqref{eq:HamiltonianSR}. We emphasize that the cc calculations need to be performed at specific experimental temperatures for the comparison with experiments due to the observed temperature-dependent splittings \cite{Zhu2019}. In the table we also give the energy differences $\Delta E_{ss}$ of Eq. \eqref{eq:EnergSS} and $\Delta E_{sr}$ of Eq. \eqref{eq:EnergySR}. Variations of the observed splittings among the five $p$-wave FRs arise from different singlet/triplet characters, thus leading to different values of $\braket{3S_z^2-S^2-1}$ in Eq. \eqref{eq:EnergSS} and $\braket{S_z}$ in \eqref{eq:EnergySR}. For the 714~G and 764~G resonances, our model predicts a small splitting between $m_N=\pm1$ components of roughly 10~mG, which is of the same size as the width of the observed loss features. Therefore, the two profiles overlap and can not be resolved under the current experimental conditions. In future experiments at lower temperatures with increased homogeneity and stability of the magnetic field this splitting might become observable.

To further justify the inclusion of \emph{sr} coupling into our model, we checked if alternative extensions of the Hamiltonian need to be included to quantitatively describe the observed splittings, e.g. the rotational Zeeman effect or the molecular anisotropy of the electron $g$-tensor with respect to the molecular axis as discussed for other molecules (see, for example, the case of O$_2$ \cite{Tischer1967}). The rotational Zeeman effect will give a splitting proportional to $m_N$ but very little dependence on the electron spin because its coupling to the rotation is too weak. Yet, the observed splitting significantly depends on the spin, which indicates that the rotational Zeeman effect does not provide a major contribution. The anisotropy of the electron $g$ tensor results also from the \emph{so} interaction. Thus, it could yield a small contribution comparable to the \emph{sr} interaction. However, when we run fitting routines including this effect, the  resulting magnitude equals about 10 \% of the free electron's g factor, which is unphysically large.
Thus, we conclude that the \emph{sr} coupling is sufficient to quantitatively describe our experimental findings.

From the cc calculations we determine the value of \emph{sr} coupling parameter as $|\gamma|=0.566(50)\times10^{-3}$. For a comparison, typical values of $\gamma$ in low-lying vibrational states of triplet diatomic molecules range from $10^{-2}$ to $10^{-3}$ \cite{Brown2003}. The sign of $\gamma$ cannot be determined from our data since changing the sign of all $m_N$, and simultaneously the sign of $\gamma$, does not change the overall quality of the fit. A positive $\gamma$ is assumed for the assignments of $m_N$ given in Fig. \ref{fig:pFRCs} and the values of $\delta^{e/cc}_{sr}$ and $\Delta E_{sr}$ listed in Table \ref{tab:pwaves}. 

Including the second-order effect, especially at short range, is essential to describing the observed \emph{ss} and \emph{sr} splittings quantitatively. The \emph{ss} interaction results from a competition between the mDDI, dominating at long range, and the second-order \emph{so} coupling, contributing significantly at short range only, due to their opposite signs in $\lambda(R)$ [$\lambda(R)=0$ at $R\approx13.9$~$a_0$]. Here $a_0$ is  the Bohr radius. In Ref. \cite{Zhu2019} we have estimated the first- and second-order contributions to the \emph{sr} coupling by assuming that only the Coulomb interactions between two atoms slightly perturb the atomic wave functions, which is a good assumption for $R>R_\mathrm{LR}\approx20$~$a_0$, the so-called LeRoy radius \cite{LeRoy1974}. In this model, the first-order effect scales as $R^{2}$ after being weighted by the nuclear wavefunction, while the second-order one scales as $R^{-3}$. However, the resulting value of $\gamma$ is much smaller than the actual one, indicating that contributions, especially the second-order ones, at $R<R_\mathrm{LR}$ are dominating. The estimate of both the first- and second-order effects at small $R$ requires ab-initio calculations \cite{Mies1996, Kotochigova2000}. The large second-order contribution in the Li-Cs system is due to the combination of a small reduced mass, giving rise to fast molecular rotation, and the large Cs \emph{so} interaction in the excited states. As a consequence, non-negligible \emph{sr} coupling is also expected for other light-heavy bi-alkali systems, like Li-Rb. 

In conclusion, by combining high-resolution Feshbach spectroscopy and cc calculations we have revealed the effect of \emph{sr} coupling giving rise to a lifting of the $|m_N| = 1$ degeneracy in $p$-wave Feshbach resonances. While this effect is well established in molecular physics, it was not anticipated to be of significance for extremely weakly bound molecular states. Our analysis highlights the importance of short-range contributions to the observed resonance splittings. By improving the magnetic-field resolution and stability as well as lowering the sample temperature, one can further increase the resolution and thus explore the structure of the electron wavefunction at short range. Such studies might reveal additional contributions of electron-electron and electron-nuclei correlations, especially for FRs occurring at very low magnetic fields. Investigations might also be extended to more deeply bound molecular states by increasing the magnetic-field strength. Triplet splittings should also be present for $p$-wave FRs in spin-polarized Fermi gases. The observed lifting of the degeneracy of the $m_N=\pm 1$ states at high magnetic fields then provides the possibility to fully control the rotational angular momentum $N$ and its projection $m_N$ in the scattering process by external magnetic fields combined with optical methods \cite{Peng2018} or appropriate trap geometries \cite{Guenter2005}. Such control might give rise to novel phases and phase transitions in $p$-wave superfluid \cite{Gurarie2005,Cheng2005}, reminiscent to the broken-axisymmetry phase in spinor BECs \cite{Murata2007}.  

\vspace{0.2cm}
\begin{acknowledgments}
		We are grateful to S. Jochim, F. Ferlaino, S. Whitlock, S. Kokkelmans and P. Fabritius for fruitful discussions. We thank M. Filzinger and E. Lippi for carefully reading the manuscript. S.H. acknowledges support by the IMPRS-QD. This work is supported by the Heidelberg Center for Quantum Dynamics, the DFG/FWF FOR2247 under WE2661/11-1, and in part by the DFG Collaborative Research Centre SFB1225 (ISOQUANT).
\end{acknowledgments}

\vspace{0.2cm}
B.Z. and S.H. contributed equally to this work.

\bibliography{Mixtures}


\end{document}